# Physics Design Considerations of Diagnostic X Beam Transport System[*]

Yu-Jiuan Chen and Arthur C. Paul, LLNL, Livermore, CA 94550, USA

*Abstract*

Diagnostic X (D-X) transport system would extract the beam from the downstream transport line of the second-axis of the Dual Axis Radiographic Hydrodynamic Test facility (DARHT-II[1]) and transport this beam to the D-X firing point via four branches of the beamline in order to provide four lines of sight for x-ray radiography. The design goal is to generate four DARHT-II-like x-ray pulses on each line of sight. In this paper, we discuss several potential beam quality degradation processes in the passive magnet lattice beamline and indicate how they constrain the D-X beamline design parameters, such as the background pressure, the pipe size, and the pipe material.

## 1 INTRODUCTION

The D-X beamlines would transport several sections of the 20 MeV, 2 kA, 2 μs long beam from the DARHT-II accelerator exit to the D-X firing point to provide four lines of sight for x-ray radiography. The DARHT-II-like x-ray pulses will be generated for each line of sight. Therefore, maintaining the beam quality in the D-X system is essential. Since the DARHT-II accelerator operates at 1 pulse per minute or less, it is also crucial to design the D-X beamline configuration requiring a minimum number of tuning shots. Modification to the DARHT-II building and the downstream transport should be minimized. Lastly, the D-X system should not preclude the future upgrade for additional lines of sight.

In this report, we discuss several beam degradation processes in the passive beamline (excluding the active components such as the kicker systems) and how they constrain the beamline design. We establish the specifications for beam emittance in Sec. 2, pipe material and size in Sec. 3, the vacuum requirement in Sec. 4. We discuss the emittance growth in a bend and the alignment specifications in Secs. 5 and 6. In Secs. 7 and 8, we discuss the design objectives on how to extract beams from the DARHT-II beamline and to minimize the tuning shots. A conclusion will be presented in Sec. 9.

## 2 EMITTANCE AND X-RAY DOSE

The internal divergence angle ($\theta = \varepsilon / a_f$) of the beam hitting the Bremsstrahlung target increase the radiation cone angle, and hence lower the forward x-ray dose. Here $\varepsilon$ is the un-normalized Lapostolle emittance, and $a_f$ is the beam radius on the target. The scaling law for the forward x-ray dose $D$ created by an electron current $I$ with a pulse length $\tau_p$ hitting a 1-mm thick tungsten plate is given by the functional form[2] of

$$D \cong 2 \times 10^{-4} I \, \tau_p \, \gamma^{2.8} \exp[-0.07\theta - 0.0103\theta^2 + 0.00087\theta^3] \,, \quad (1)$$

with $I$ in kA, $\tau_p$ in ns, $\theta$ in degrees, and $D$ in Roentgen at 1 m. The DARHT-II final emittance specification is 1500 π mm-mr. To achieve the DARHT-II x-ray dose within ±5 % with a 2.1 mm diameter spot size, the final emittance for the D-X should be less than 1875 π mm-mr.

## 3 CONDUCTIVITY AND PIPE SIZE

There are several transport issues regarding the choice of beam pipe size and material, especially, conductivity of the pipe material. The finite conductivity of the pipe wall can cause both transverse[3] and longitudinal[4] resistive wall instabilities and beam energy loss.

### 3.1 Transverse Resistive Wall Instability

The transverse resistive wall instability arises from the diffusion of the return current into the wall due to its finite conductivity. The reduction in the magnetic forces of the dissipating return current acts as frictional forces on the particles. The friction forces enhance the slow wave of the transverse beam displacement over the length of the beam path and lead to beam loss and large time integrated spot sizes on the targets.

For a continuous focusing system, the instability grows approximately as $\exp[1.5(z/L_{tr})^{2/3}]$, where $L_{tr}$ is the instability's characteristic growth length[3] given as

$$L_{tr} = \frac{2\gamma \beta I_o}{I} \sqrt{\frac{\pi\sigma}{\tau} \frac{k_\beta b^3}{c}} \,, \quad (2)$$

where $k_\beta$ is the space charge suppressed betatron wavenumber, $I_o$ is the Aflvén current (~ 17 kA), $\sigma$ and $b$ are the beampipe's conductivity and radius. The characteristic growth length ($l_{tr}$) in a drift region is given as

$$l_{tr} = \sqrt{\frac{\gamma \beta I_o}{2I}} \sqrt[4]{\frac{\pi\sigma}{\tau}} \sqrt{\frac{b^3}{c}} \,. \quad (3)$$

Comparing the equations above, we observe that $l_{tr} = (L_{tr}/k_\beta)^{1/2}/2$. Chopping the 2 μs long pulse to shorter pulses soon, using a large, highly conductive beampipe, and transporting beam with relatively strong focusing fields can reduce the instability growth.

Two pipe materials are examined for the instability growth: 304 stainless steel at 68 F°, and 6061-TO Al with at 68 F°. Based on the growth lengths given in Table 1, we have chosen the beampipe to be 8 cm radius, 6061-TO Al. Note that the beams in most of the D-X beamline are short pulses (~100 ns) extending over 2 μs instead of the full 2 μs long pulses, and hence, the true growth lengths are expected to be longer than those in Table 1. We have also limited the maximum length for each beamline branch to 200 m, and that for individual drift sections to 10 m. Since the nominal D-X beam radius is 1 cm, the chance of having the beam melt the Al pipe wall should be small.

| Pipe material | Conductivity (sec$^{-1}$) | Overall System's Growth Length | Growth Length in Drift Region |
|---|---|---|---|
| 304 SS | $1.25 \times 10^{16}$ | 51.2 m | 6.4 m |
| 6061-TO Al | $2.368 \times 10^{17}$ | 222.8 m | 13.3 m |

Table 1 The resistive wall instability's growth length for a 2 µs long D-X beam. The pipe radius is 8 cm.

Using a 13 cm radius stainless steel pipe also gives the same growth length. However, the transport system would be 60 % longer. Furthermore, the accuracy of the beam position monitors needed for a larger pipe would lead to a larger uncorrectable centroid displacement which provides the free energy for emittance growth.

### 3.2 Parasitic Energy Loss

When the beam is traveling down a pipe, the return current flows in the resistive pipe wall in the opposite direction and creates a longitudinal voltage drop along the wall. For a 2 kA beam pulse with a 5 ns rise time traveling for 100 m in a 8 cm radius Al pipe, the energy loss at 50% of the peak current is only about 1 keV, i.e., 0.005 % of the beam energy.

### 3.3 Longitudinal Resistive Wall Instability

The ohmic losses due to the image currents flowing in the resistive wall coupled with perturbations in the current's line density lead to longitudinal resistive wall instability. Since the ohmic losses are negligible for the D-X transport, the beam needs to travel a long distance before accumulating enough perturbation in the line density from bunching or debunching. Therefore, the longitudinal resistive wall instability should not a concern. Without any Landau damping, the growth length is 98 km for the instability at 1 GHz and 219 km for the instability at 200 MHz.

## 4 VACUUM REQUIREMENT

Collisional ionization of the background gas may affect the beam propagation in a substantial way. Secondary electrons created in the collisions can be expelled by the beam electrons' space charge fields, while the ions remain trapped in the potential well of the electron beam. Those ions forming an ion channel can provide the background gas focusing forces and cause the ion hose instability[5-9]. The background gas can also cause an emittance growth by scattering the beam electrons.

### 4.1 Background Gas Effects

The accumulated difference in the phase advance between the beam head and the beam tail over a distance L due to the ion focusing effects of the background gas can be approximated as

$$\Delta\phi \approx \sqrt{\frac{I}{\gamma\beta^3 I_o} \tau_p[\text{ns}] P[\text{torr}] \frac{L}{a}} \quad , \quad (4)$$

A large difference in the phase advance causes beam envelope oscillations within a pulse, and hence, increase the time integrated spot size. Therefore, to achieve a small time integrated spot size, $\Delta\phi$ should be much less than $2\pi$. The pressure in the D-X needs to be much less than $5 \times 10^{-8}$ torr for a 2 µs beam in a 120m long beamline. However, the electron beam is 2 µs long only while it is being extracted from the DARHT-II beamline. It then will be chopped into four shorter pulses with the longest pulse being about 500 ns wide. The subsequent kickers will chop these four pulses further into even shorter pulses. Most of the ions trapped by the electron beam's space charge potential well would escape to the wall quickly during the pulse separation time. Therefore, a vacuum of $\sim 10^{-8}$ torr should be sufficient.

### 4.2 Ion Hose Instability

The ion-hose instability occurs when the electron beam is not collinear with the ion channel. The instability grows with time and distance. It grows the fastest when the beam travels in a preformed offset, constant-strength ion channel with all the ions oscillating at the same frequency. The maximum number of e-folding growth[6] ($\Gamma$) for an emittance dominated Bennett beam at its equilibrium travel a distance $L$ is given as

$$\Gamma \cong \frac{3 I f_e L}{\gamma \beta^3 I_o \varepsilon} \quad . \quad (5)$$

Let $d$ be the initial separation between the beam and the ion channel, and $\lambda$ be the instability's transverse beam displacement normalized to the beam radius. We now have $\lambda = (d/a) e^{\Gamma}$. Rearranging Eq. (5) yields to the pressure requirement as

$$P[\text{torr}] \cong \frac{\gamma\beta^3}{3\tau_p[\text{ns}]} \frac{I_o}{I} \varepsilon[\text{cm-rad}] \ln\left(\lambda \frac{a}{d}\right) \quad . \quad (6)$$

Assume that the initial beam-channel separation is 0.5 mm, for a 2 µs long, 1 cm radius beam. To achieve a transverse beam motion at a x-ray target less than 10% of the beam size, $P \leq 1 \times 10^{-8}$ torr. A spread in ion's mass or oscillation frequency damps the instability growth[7]. The detuning of the ion oscillating frequency due to the beam envelope changes either by design or by mismatch along the beamline also suppress the growth[8]. Furthermore, the recent studies[8, 9] suggest that its saturation amplitude is always small compared with the beam radius. All these may lead to a more relaxed vacuum requirement.

### 4.3 Background Gas Scattering

Scattering in a background gas can cause emittance growth. For a 20 MeV beam traveling 120 m in $10^{-8}$ torr of oxygen, the total emittance growth caused by background gas scattering is $1.38 \times 10^{-3}$ π mm-mr which is insignificant compared with the nominal 1000 π mm-mr emittance exiting the DARHT-II accelerator.

## 5 EMITTANCE GROWTH IN BENDS

A typical D-X beamline consists of 4 bending systems and two active kickers. To achieve the emittance specification at the D-X targets, the relative emittance growth for each bending system is limited to 5 %. According to Ref. 10, the emittance growth arising from nonlinear fields in the bend geometry is negligible for the D-X. Most of the emittance growth would come from the nonuniform image charge and current distributions due to the geometric difference between the cross-sections of the beam and the pipe[11]. To minimize the emittance growth, the shape of the beam pipe needs to be the same as the beam. Typically, the shape of the beam cross-section in a bending system, consisting of several dipole magnets and quadrupoles, varies from magnet to magnet. A nominal D-X bending system is about 10 m in length. Making the pipe cross-section similar to the beam cross-section along the bend will be difficult and costly. For practical reasons, the pipe cross-section is round through out the D-X beamline except where the beampipe is split. The D-X transport tune is chosen to make the beam as round as possible and its size much smaller than the pipe size so that the nonlinear image force effects are small.

## 6 MISALIGNMENT

Coupling of chromatic aberration of a misaligned transport system and the time dependent energy variation within a current pulse leads to a time varying transverse motion called corkscrew motion[12, 13]. When the difference in the total phase advance ($\phi_{tot}$) of the beam transport through the entire system is small compared with unity, the corkscrew amplitude for this transport system with randomly misaligned magnets is given as

$$|\eta| \approx n^{3/2} |\rho_{rms}| \left(\frac{\Delta\gamma}{\gamma}\right)\phi$$
$$\approx \sqrt{n} |\rho_{rms}| \left(\frac{\Delta\gamma}{\gamma}\right)\phi_{tot} \quad , \quad (7)$$

where $\phi$ is the phase advance after traveling through a single misaligned magnet. The quantity $\rho_{rms}$ is the r.m.s. value of the electron gyro-radii of the reference beam slice resulting from $n$ randomly misaligned magnets.

Assume $|\rho_{rms}| \sim \Delta\theta_{rms} l \sim \Delta_{rms}$, where $\Delta\theta_{rms}$ is the r.m.s. value of the magnet tilts, $\Delta_{rms}$ is the r.m.s. value of the magnet offsets, and $l$ is the magnet length (nominally 20 cm). The amplitude of the time varying transverse beam motion on the DARHT-II target should be less than 10% of the beam radius. The beam typically has to travel through approximately 100 more magnets to reach its D-X converter target than to reach the DARHT-II target. Let us assume that the corkscrew amplitude arising from misalignment of those 100 magnets alone is also one tenth of the beam radius. The corkscrew amplitudes caused by random tilts and offsets generally add in quadrature. The net corkscrew amplitude, including the contribution from the misaligned DARHT-II magnets, at a D-X target will be about 14 % of the beam radius. For the worst case, the time integrated D-X spot size is now 14 % larger versus 10 % larger for the time integrated DARHT-II spot size if the same focusing field is used for both systems. We can easily compensate for this difference by increasing the final focusing field at the D-X target area by 4 %. The energy variation within the pulse is less than or equal to ± 0.5 %. The total phase advance on a D-X beamline is nominally ~14π. We assume that the random tilts and the random offsets contribute to the corkscrew motion equally. According to Eq. (7), the r.m.s. tilt has to be less than 0.8 mrad, and the r.m.s. offset has to be less than 0.16 mm if no dynamic steering would be implemented to control the corkscrew motion. These specifications are very similar to the DARHT-II alignment specifications: 1.95 mrad for the 3 σ in tilts and 0.45 mm for the 3 σ in offsets. Our experience on dynamically controlling the corkscrew motion on ETA-II[13] and FXR[14] and the simulations of corkscrew control on DARHT-II indicate that corkscrew motion on D-X should not be a concern.

## 7 BEAM EXTRACTION AND MODULARITY OF THE BEAMLINE

The first design task in the Diagnostic X beamline is the beam extraction from DARHTII. Due to the space constraint, the beam will be extracted before it enters the DARHT-II kicker. The beam will be 2 μs long with an unknown amount of beam head current. To prevent the beam from drilling a hole through the D-X Al beampipe, the extraction line should coexist with the DARHT-II shuttle dump so that the dump can be used to characterize the beam and to tune the DARHT-II accelerator before bending the beam into the D-X transport system.

In order to minimize the number of shots needed for tuning the D-X beamline, the beamline will consist of several "identical" modules, such as 10°, 45° and 90° achromatic bend systems, 90° kicker achromatic bend systems, etc. A matching section precedes each module. Operationally, we would tune each type of module once, and then set all the identical module's magnets at the same tune. We then would only tune individual matching sections to match beams into the following modules.

## 8 CONCLUSIONS

Many transport concerns have been discussed in this paper. Based on our current knowledge of those potentially degrading mechanisms, we believe that they can be suppressed.